\documentclass{jpp}
\usepackage{graphicx}
\usepackage[hidelinks]{hyperref}

\usepackage[table]{xcolor}

\usepackage{amsfonts}
\usepackage{mathrsfs}

\usepackage[utf8]{inputenc}
\usepackage[T1]{fontenc}
\usepackage{amsmath}

\newcommand{\bv}{\mathbf{v}}
\newcommand{\bE}{\mathbf{E}}
\newcommand{\bB}{\mathbf{B}}

\newcommand{\pa}[2]{\frac{\partial #1}{\partial #2}}
\newcommand{\xv}{z}

\newcommand{\D}{\mathrm{d}}

\shorttitle{Loss-Cone Stabilization in Rotating Mirrors}

\title{Loss-Cone Stabilization in Rotating Mirrors: Thresholds and Thermodynamics}

\author{E. J. Kolmes
  \corresp{\email{ekolmes@princeton.edu}},
  I. E. Ochs,
 \and N. J. Fisch}

\affiliation{Department of Astrophysical Sciences, Princeton University, Princeton, NJ 08544, USA}

\begin{document}

\maketitle

\begin{abstract}
In the limit of sufficiently fast rotation, rotating mirror traps are known to be stable against the loss-cone modes associated with conventional (non-rotating) mirrors. 
This paper calculates how quickly a mirror configuration must rotate in order for several of these modes to be stabilized (in particular, the high-frequency convective loss cone, drift cyclotron loss cone, and Dory-Guest-Harris modes). 
Commonalities in the stabilization conditions for these modes then motivate a modified formulation of the Gardner free energy and diffusively accessible free energy to be used for systems in which the important modes have wave vectors that are orthogonal or nearly orthogonal to the magnetic field, as well as a modification to include the effects of a loss region in phase space.
\end{abstract}

\section{Introduction}

The rotating magnetic mirror is a plasma confinement concept with a number of advantages. 
This class of device modifies a typical mirror configuration by adding large, approximately-radial electric fields perpendicular to the magnetic fields, thereby producing fast rotation \citep{Lehnert1971, Abdrashitov1991, Ellis2001, Teodorescu2010}. 
One purpose of this rotation is to improve longitudinal confinement; the projection of the centrifugal force along the magnetic field lines is everywhere oriented toward the middle of the trap, so that the centrifugal potential acts to trap particles. 
Another purpose is to improve the stability of the device. 

This stability improvement happens through two main mechanisms. 
If the rotation is sheared, the shear flow can act to suppress a wide range of instabilities, perhaps even reducing cross-field transport to classical levels \citep{Piterskii1995, Hassam1999, Cho2005, Maggs2007, Beklemishev2010}. 
The other mechanism -- and the focus of this article -- is that rotation modifies the phase-space loss cone in such a way as to reduce or even eliminate loss-cone instabilities. 

Loss-cone instabilities are modes that draw free energy from the ion population inversion that exists as a result of the loss conditions in the device \citep{Post1987}. 
In the absence of any electrostatic potential, a non-rotating mirror has a loss region that forms a cone in phase space. 
In the presence of a confining potential, that cone is ``lifted'' to form a hyperboloid of revolution. 
This is shown schematically in Figure~\ref{fig:cartoonLossCones}, and will be discussed in greater detail in Section~\ref{sec:models}. 
A confining potential could be centrifugal or electrostatic. 
In general, it may be both; virtually all low-collisionality mirror systems have some electrostatic potential equalizing the ion and electron loss rates. 
Fundamentally, none of the underlying physics discussed in this paper is specific to one kind of potential or another. 

Rotation (or any other confining potential) moves the region with the population inversion to higher velocities, away from the most-populated part of phase space. 
In the limit of very fast rotation, the hyperboloid can be lifted away from all but the highest-energy superthermal particles, and it is clear that the instability must be suppressed. 

What is less clear is the threshold at which this suppression will take place. 
Indeed, though it is common in the rotating-mirror literature to assert that these modes ought to be stabilized, calculations of the actual threshold and details of this stabilization are few and far between. 
To the authors' knowledge, there have been two such studies: one by \citet{Volosov1969}, and the other by \citet{Turikov1973}. 
Both address only a single loss-cone instability: the high-frequency convective loss-cone (HFCLC) mode. 

The first part of this paper seeks to provide a more detailed treatment of the stabilizing effects on loss-cone modes, including the high-frequency convective loss cone mode (HFCLC), the drift cyclotron loss cone mode (DCLC), and the Dory-Guest-Harris mode (DGH) \citep{Rosenbluth1965, Dory1965, Post1966, Post1987, Kotelnikov2017}. 
The analysis includes two different models for the effects of rotation on the distribution function; these models show qualitatively similar trends but the details of their behavior differ significantly. 
It also includes numerical validation. 

Over the course of that analysis, we will find that these three modes share a sufficient condition for stabilization. 
Moreover, all of these modes share an intuitive physical picture in which the mode is driven (at least in part) by a population inversion in phase space, so that energy can be released by rearrangements of the distribution that relax that inversion. 
We will also find a surprise: that the HFCLC and DCLC both appear to be harder to stabilize when the mirror ratio $R$ is larger. 
This kind of $R$ dependence was also observed by \citet{Turikov1973}, who concluded that it must come from a deficiency of the model distribution function used to calculate the stability threshold. 
We will explore the stabilization conditions with several different models and find that the same trend appears in all of them. 

Taken together, these three factors motivate the second part of this paper, which seeks to explain these shared characteristics in terms of the theory of plasma free energy. 
It turns out that the theories of Gardner restacking and diffusive exchange can explain these stabilization thresholds, including their surprising $R$ dependence, but only if the free energy is modified to incorporate constraints associated with the flute-like structure of the relevant modes. 

This paper is organized as follows. 
Section~\ref{sec:models} discusses different models for the effects of rotation on the plasma distribution. 
Sections~\ref{sec:HFCLC}, \ref{sec:DCLC}, and \ref{sec:DGH} calculate stabilization thresholds for the HFCLC, DCLC, and DGH modes, respectively. 
Section~\ref{sec:fluteRearrangements} describes how the existing theories of free energy can be modified to take into account the constraint that rearrangements are being produced by flute-like modes. 
Section~\ref{sec:lossConeRearrangements} describes the further modifications that are necessary in order to properly account for the existence of a loss cone. 
Section~\ref{sec:conclusion} summarizes and discusses these results. 

\section{Loss Cones and Analytic Models} \label{sec:models}

\begin{figure}
    \centering
    \includegraphics[width=\linewidth]{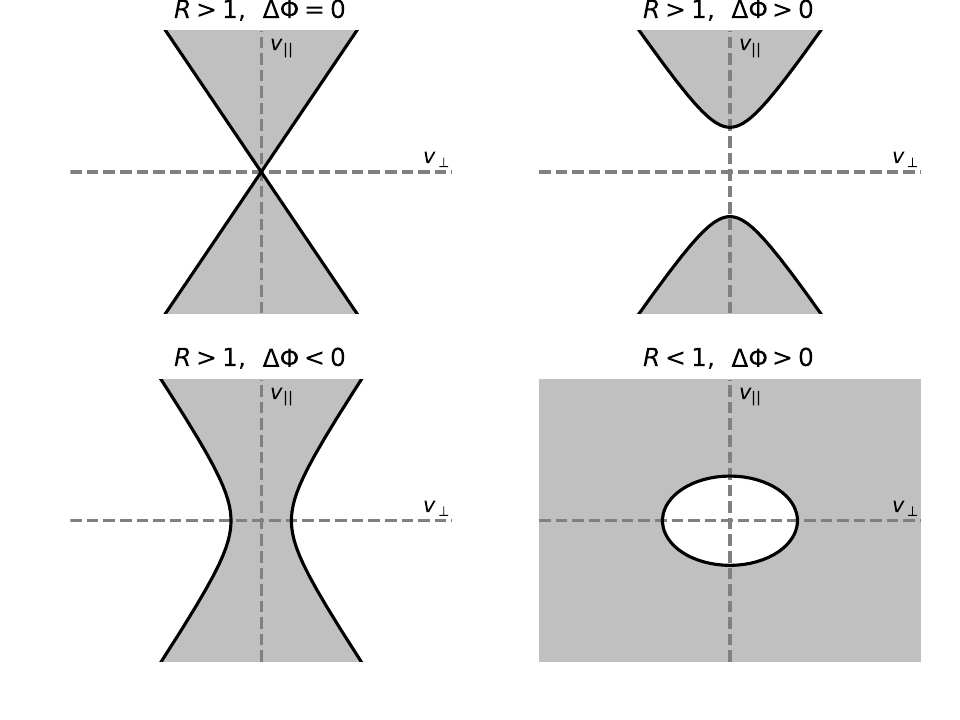}
    \caption{Trapped (unshaded) and untrapped (shaded) regions of velocity space for different signs of $\Delta \Phi$ and $R-1$, where $\Delta \Phi$ is the axial potential energy drop (including centrifugal and electrostatic effects) and $R$ is the mirror ratio.}
    \label{fig:cartoonLossCones}
\end{figure}

Before discussing the loss-cone instabilities themselves, it is important to discuss the structures of the loss cones themselves. 
Consider a system with magnetic field strength $B_\text{mid}$ at the midplane and field strength $B_\text{end}$ at the mirror ends. 
Let $R$ denote the mirror ratio, which is defined by 
\begin{gather}
R \doteq \frac{B_\text{end}}{B_\text{mid}} \, . 
\end{gather}
Let $\bv$ denote the velocity of a particle and $m$ denote its mass. 
Define $B \doteq |\bB|$, $\hat b \doteq \bB / B$, $v_{||} \doteq \hat b \cdot \bv$, $\bv_\perp \doteq \bv - v_{||} \hat b$, and $v_\perp \doteq |\bv_\perp|$. 
Furthermore, let $\mu$ denote the first adiabatic invariant: 
\begin{gather}
\mu \doteq \frac{m v_\perp^2}{2 B} \, . 
\end{gather}
In the limit in which collisions can be neglected, and in which the rotation frequency is small compared with the gyrofrequency \citep{Volosov1969, Thyagaraja2009}, this $\mu$ is conserved. 
For a particle subject to some potential $\Phi$ (including the electrostatic potential and, in a rotating system, the centrifugal potential) the conservation of energy can be written as 
\begin{gather}
\frac{1}{2} m v_{||}^2 + \mu B + \Phi = \text{constant}. 
\end{gather}
Then if $\Delta \Phi$ is the difference in $\Phi$ between the point of maximum $B$ and the midplane (and assuming that $\Phi$ does not have interior extrema away from the midplane) the condition for a particle to be trapped is 
\begin{gather}
v_{||}^2 \leq (R-1) v_\perp^2 + \frac{2 \Delta \Phi}{m} \, . \label{eqn:CartesianTrappingCondition}
\end{gather}
This condition is written in terms of the particle's midplane velocity. 
If Eq.~(\ref{eqn:CartesianTrappingCondition}) is rewritten in terms of spherical  $(v,\theta,\varphi)$ velocity coordinates, this becomes 
\begin{gather}
\sin^2 \theta \geq \frac{1}{R} \bigg( 1 - \frac{2 \Delta \Phi}{m v^2} \bigg), \label{eqn:angularTrappingCondition}
\end{gather}
which is the well-known generalization of the mirror loss cone to the case with a potential $\Phi$ \citep{Pastukhov1987}. 
The trapping regions defined by Eq.~(\ref{eqn:angularTrappingCondition}) are shown in Figure~\ref{fig:cartoonLossCones}. 
When $\Delta \Phi = 0$, the trapped region is a cone in phase space. 
When $\Delta \Phi \neq 0$ and $R > 1$, the trapped region becomes a hyperboloid of revolution. 
Finally, when $\Delta \Phi > 0$ and $R < 1$, the trapped region is ellipsoidal. 
This last case is not a focus of the present paper but it has advantages in certain scenarios \citep{Volosov2006, Mlodik2023, Munirov2023}. 
Other cases with $R < 1$ are not shown because these cases do not have a trapped region. 

Over the course of this paper, it will often be important to write down a model for a distribution function consistent with this trapping condition for a given $R$ and $\Delta \Phi$ (neglecting, for simplicity, any spatial variation within the trapped region). 
Perhaps the simplest choice is a truncated Maxwellian: 
\begin{gather}
f_T(\bv) = A e^{- m v^2 / 2 T} \Theta \bigg[ \frac{2 \Delta \Phi}{m} + (R-1) v_\perp^2 - v_{||}^2 \bigg]. \label{eqn:truncatedMaxwellian}
\end{gather}
Here $\Theta$ is the Heaviside step function and the normalization constant $A$ is chosen such that 
\begin{gather}
\int \D^3 \bv \, f(\bv) = 1. 
\end{gather}
This model is simple and easy to understand, but does have the disadvantage of having (arguably unphysically) sharp boundaries around the trapping region. 

Another choice, essentially equivalent to one used by \citet{Volosov1969} and \citet{Turikov1973}, is 
\begin{gather}
f_S(\bv) = A \sqrt{ \frac{2 \Delta \Phi}{m} + (R-1) v_\perp^2 - v_{||}^2} \, e^{-m v_\perp^2 / 2 T} \Theta \bigg[ \frac{2 \Delta \Phi}{m} + (R-1) v_\perp^2 - v_{||}^2 \bigg] . \label{eqn:fVolosov}
\end{gather}
This model has a smoother transition to the loss cone. 
Its original use by \citet{Volosov1969} was motivated by the fact that, when $\Delta \Phi = 0$ and $R = 2$, it reduces to an analytic model of a non-rotating mirror used by \citet{Rosenbluth1965}. 
However, it comes with a major disadvantage: it does not reduce to a Maxwellian when $\Delta \Phi$ is large. 
This suggests that $f_S$ might be a reasonable model in the limit of slow rotation, but that it is unphysical for larger $\Delta \Phi$. 

For practical purposes, as well as to facilitate easier translation between our results and those elsewhere in the literature, it may be helpful to explicitly show the mapping between $\Delta \Phi$ and the rotational Mach number. 
Suppose $\Delta \Phi$ is entirely due to a centrifugal potential (that is, suppose electrostatic effects can be neglected). 
Then if the angular frequency of rotation $\Omega$ is constant along a flux surface, and if $r_\text{min}$ and $r_\text{max}$ are the minimum and maximum radii along the flux surface, a particle of mass $m_i$ sees 
\begin{gather}
\Delta \Phi = \frac{1}{2} m_i (r_\text{max}^2-r_\text{min}^2) \Omega^2 . 
\end{gather}
If $R = r_\text{max}^2 / r_\text{min}^2$, then this can be rewritten in terms of the mirror ratio $R$ as 
\begin{gather}
\Delta \Phi = \frac{R-1}{2 R} m_i r_\text{max}^2 \Omega^2 . 
\end{gather}
Then if the rotational Mach number is defined as 
\begin{gather}
\text{Ma} \doteq \sqrt{ \frac{m_i r_\text{max}^2 \Omega^2}{2 T} }
\end{gather}
for a temperature $T$, we have 
\begin{gather}
\frac{\Delta \Phi}{T} = \frac{R-1}{R} \, \text{Ma}^2. \label{eqn:phiMa}
\end{gather}
Of course, Eq.~(\ref{eqn:phiMa}) can (and, for most real devices, will) be modified by electrostatic potentials, which need to be included in $\Delta \Phi$ alongside the centrifugal potential. 
Quasineutrality generally requires the combined electrostatic and centrifugal potentials for electrons and ions to be roughly equal. 
In the case of singly-charged ions and equal ion and electron temperatures, this leads to an electrostatic potential with a magnitude half that of the centrifugal potential, equalizing the species' confining potentials by trapping the electrons and reducing the trapping of the ions. 
It will also be modified in cases where $R \neq r_\text{max}^2 / r_\text{min}^2$. 
This can happen if the plasma occupies a volume with an annular rather than circular cross-section. 
It can also result from plasma diamagnetism. 
As such, Eq.~(\ref{eqn:phiMa}) should be understood as a reasonable scaling but not as a precise mapping for all devices. 
In any case, an important takeaway from Eq.~(\ref{eqn:phiMa}) is that when considering stability thresholds described in the subsequent sections of this paper, one should be aware that $\Delta \Phi$ can be understood largely as a proxy for the rotation speed but that the mapping between $\Delta \Phi$ and $\text{Ma}$ does also depend on the mirror ratio $R$. 

\section{High-Frequency Convective Loss Cone Instability} \label{sec:HFCLC}

The first loss-cone mode to be considered is the high-frequency convective loss cone (HFCLC) instability. 
In the regime in which it is unstable, the HFCLC is characterized by traveling waves that grow as they propagate. 
As we will see, the typical wavelengths are on the scale of the ion Debye length. 

The HFCLC is derived using a number of assumptions. 
First, the ion plasma frequency $\omega_{pi}$ and ion cyclotron frequency $\omega_{ci}$ satisfy $\omega_{pi}\gg\omega_{ci}$. 
This is a statement about the density regime of interest. 
Second, the parallel and perpendicular components of the wavenumber $\mathbf{k}$ satisfy $k_{||} \ll k_\perp$. 
This constraint is often imposed in mirror-type systems due to the large electron mobility. 
Third, the mode frequency $\omega$ satisfies $\omega_{ci} \ll \omega \ll \omega_{ce}$. 
This means that the plasma response can be calculated under the assumption that the Larmor gyration of the ions can be ignored but that the electron motion is almost entirely along field lines \citep{Post1987}. 
Then, not taking into account the stabilizing effects of finite electron temperature, the electrostatic dispersion relation can be written as \citep{Rosenbluth1965, Post1966, Post1987}
\begin{gather}
1 + \frac{\omega_{pe}^2}{\omega_{ce}^2} = \frac{\omega_{pe}^2}{\omega^2} \frac{k_{||}^2}{k^2} + \frac{\omega_{pi}^2}{k^2 \bar{v}_i^2} F \bigg( \frac{\omega}{k_\perp \bar{v}_i} \bigg), 
\end{gather}
where $\bar{v}_i$ is the ion thermal velocity and 
\begin{gather}
F(y) \doteq - 2 \int_0^\infty \bigg( 1 - \frac{x}{y^2} \bigg)^{-1/2} \frac{\partial \psi}{\partial x} \, \D x \label{eqn:Fy} \\
\psi\bigg( \frac{v_\perp^2}{\bar v_i^2} \bigg) \doteq \bar\psi \int_{-\infty}^{+\infty} f(v_\perp^2,v_{||}^2) \, \D v_{||}, \label{eqn:psiDefinition}
\end{gather}
in which $f$ is the ion distribution function, $x = v_\perp^2 / \bar v_i^2$, and the normalization constant $\bar \psi$ is chosen such that 
\begin{gather}
\int_0^\infty \psi(x) \, \D x = 1. 
\end{gather}
\citet{Post1966} showed that the most unstable $\mathbf{k}$ satisfies 
\begin{align}
| \text{Im}(k_{||}) |_\text{max} \approx \frac{1}{2} \sqrt{\frac{m_e}{m_i}} \frac{\omega_{pi}}{\bar{v}_i} \bigg( 1 + \frac{\omega_{pe}^2}{\omega_{ce}^2} \bigg)^{-1/2} \big[ - y \text{Im} F(y) \big]_\text{max}, 
\end{align}
where $m_e / m_i$ is the electron-ion mass ratio, $\omega_{ps}$ is the plasma frequency of species $s$, $\omega_{cs}$ is the cyclotron frequency of species $s$, and $-y \text{Im} F(y)$ is maximized over $y$. 

Note that 
\begin{gather}
-y \text{Im} F(y) = 2 y^2 \int_{y^2}^\infty \big( x - y^2 \big)^{-1/2} \frac{\partial \psi}{\partial x} \, \D x . \label{eqn:yImFy}
\end{gather}
It is clear from Eq.~(\ref{eqn:yImFy}) that the mode cannot be unstable if $\psi$ is a monotonically decreasing function of $x$.\footnote{This monotonicity condition can also be derived using a Nyquist-Penrose approach, and is equivalent to the Penrose monotonicity condition given appropriate assumptions on the symmetry of the distribution function and orientation of the wavenumber \citep{Penrose1960}. The Nyquist-Penrose approach for this problem is discussed further in \citet{Rosenbluth1965}. }
Similarly, the mode cannot be unstable if there is no choice of $y$ for which $y \text{Im}F(y)$ is negative. 
The second condition turns out to be more easily satisfied than the first. 
It is also possible to define a third, even more easily satisfied condition by requiring that the mode must have $\text{Im}(\omega) > \omega_{ci}$, since this is an additional requirement of the HFCLC \citep{Mikhailovskii}. 
This leads to a requirement that $-y \text{Im} F(y)$ be larger than a finite threshold rather than simply that it be positive. 
This was the approach of Turikov. 
This third condition is a density-dependent correction to the second, and depends on the value of $\omega_{pi} / k \bar{v}_i$. 
In the limit where $\omega_{pi} \gg k \bar{v}$, the second and third conditions are equivalent. 
For the sake of simplicity (and a smaller parameter space), we will focus on the first and second sufficient conditions for stability, which can be summarized as follows: 

\textit{Perpendicular Monotonicity Condition:}
\begin{gather}
\frac{\partial \psi}{\partial x} \leq 0 \quad \forall x \geq 0. \label{eqn:monotonicityCondition}
\end{gather}

\textit{HFCLC Integral Condition:}
\begin{gather}
\int_{y^2}^\infty \big( x - y^2 \big)^{-1/2} \frac{\partial \psi}{\partial x} \, \D x \leq 0 \quad \forall y > 0. \label{eqn:hfclc}
\end{gather}
Either condition serves independently as a sufficient condition for HFCLC stability; the distributions that satisfy the former condition are a subset of the distributions that satisfy the latter. 

\begin{figure}
    \centering
    \includegraphics[trim={2.7cm, 0, 2.7cm, 0}, clip, width=\linewidth]{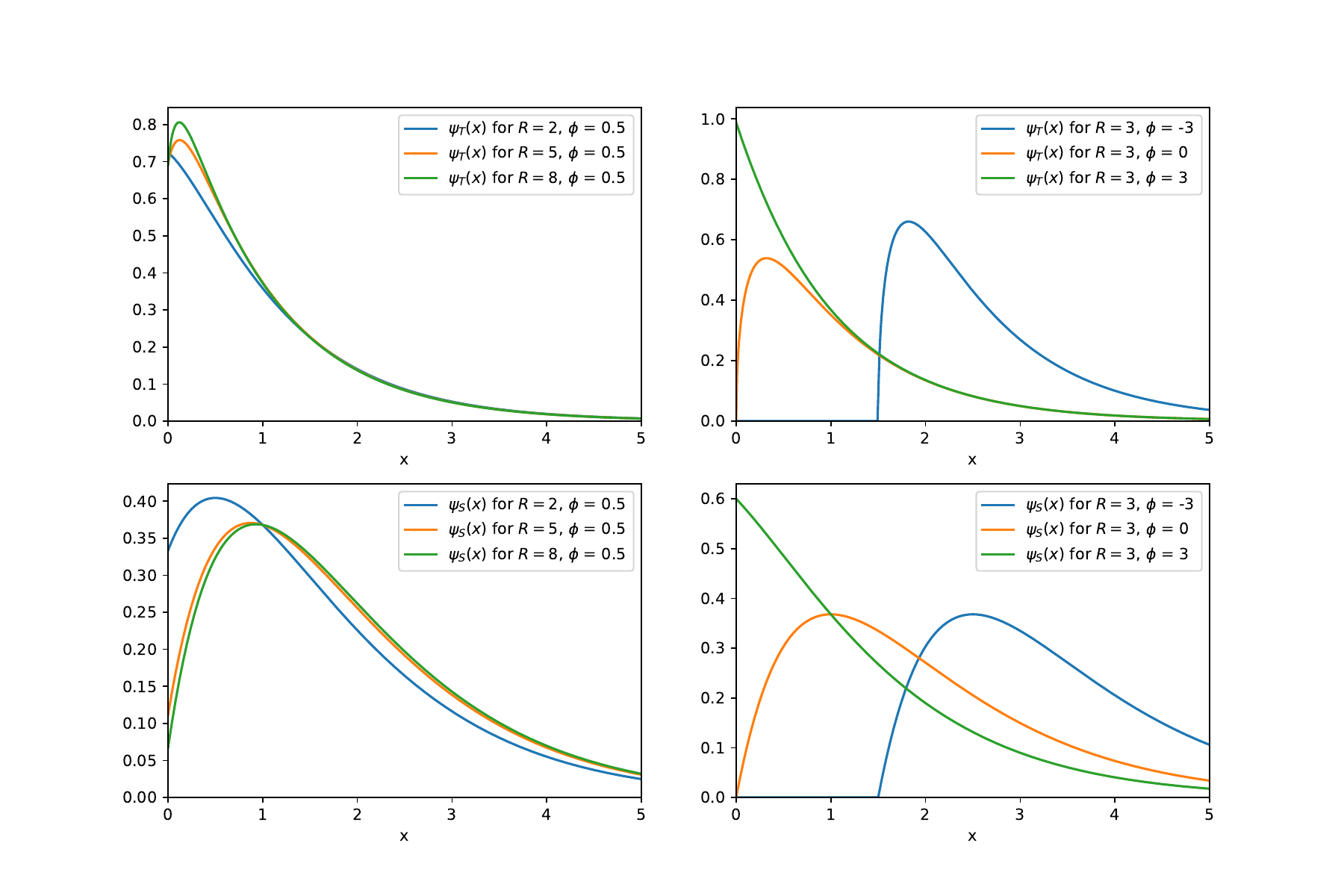}
    \caption{The projected perpendicular-energy distribution functions $\psi_T$ and $\psi_S$ for several choices of $R$ and $\phi$. }
    \label{fig:psiPlots}
\end{figure}

These conditions can be evaluated for each of the distributions discussed in Section~\ref{sec:models}. 
Going forward, it will be convenient to work with the dimensionless confinement potential 
\begin{gather}
\phi \doteq \frac{\Delta \Phi}{T} \, . 
\end{gather}
When $\phi > 0$, the potential is confining. 
First, integrating $f_T$ from Eq.~(\ref{eqn:truncatedMaxwellian}), we find that the corresponding $\psi$ is 
\begin{align}
\psi_T(x) &= \psi_0 e^{-x} \text{erf} \bigg[ \sqrt{ (R-1) x + \phi } \bigg] \Theta \bigg[ (R-1) x + \phi \bigg], 
\end{align}
where $\Theta$ is again the Heaviside step function. 
The normalization constant $\psi_0$ is 
\begin{align}
\psi_0^{-1} = \text{erf} \bigg[ \sqrt{\phi} \bigg] + e^{\phi/(R-1)} \text{erfc} \bigg[ \sqrt{ \frac{R \phi}{R-1} } \bigg] \quad \text{for } R>1, \phi \geq 0
\end{align}
and 
\begin{align}
\psi_0^{-1} = e^{\phi / (R-1)} \sqrt{\frac{R-1}{R}} \quad \text{for } R>1, \phi < 0. 
\end{align}
Integrating $f_S$ from Eq.~(\ref{eqn:fVolosov}) with respect to $v_{||}$, 
\begin{align}
\psi_S(x) &= \psi_0 \big[ (R-1) x + \phi \big] \, e^{-x} \, \Theta \bigg[ (R-1) x + \phi \bigg] 
\end{align}
where 
\begin{align}
\psi_0^{-1} = \phi + R - 1 \quad \text{for } R>1, \phi \geq 0
\end{align}
and 
\begin{align}
\psi_0^{-1} = (R-1) e^{\phi/(R-1)} \quad \text{for } R>1, \phi < 0. 
\end{align}
$\psi_T$ and $\psi_S$ are plotted for several choices of $R$ and $\phi$ in Figure~\ref{fig:psiPlots}. 

The first sufficient condition for stabilization -- that is, the monotonicity of $\psi$ -- is relatively easy to check. 
Assuming $R > 1$, neither $\psi_T$ nor $\psi_S$ will meet this condition when $\phi < 0$. 
$\psi_T$ satisfies the perpendicular monotonicity condition when 
\begin{gather}
\sqrt{\pi \phi} \, e^\phi \, \text{erf}\sqrt{\phi} \geq R-1 . 
\end{gather}
$\psi_S$ satisfies the same monotonicity condition when 
\begin{gather}
\phi \geq R - 1. 
\end{gather}
Note that in both cases, higher $R$ pushes the stability threshold to higher $\phi$. 

\begin{figure}
    \centering
    \includegraphics[trim={1.5cm, 0, 1.5cm, 0}, clip, width=\linewidth]{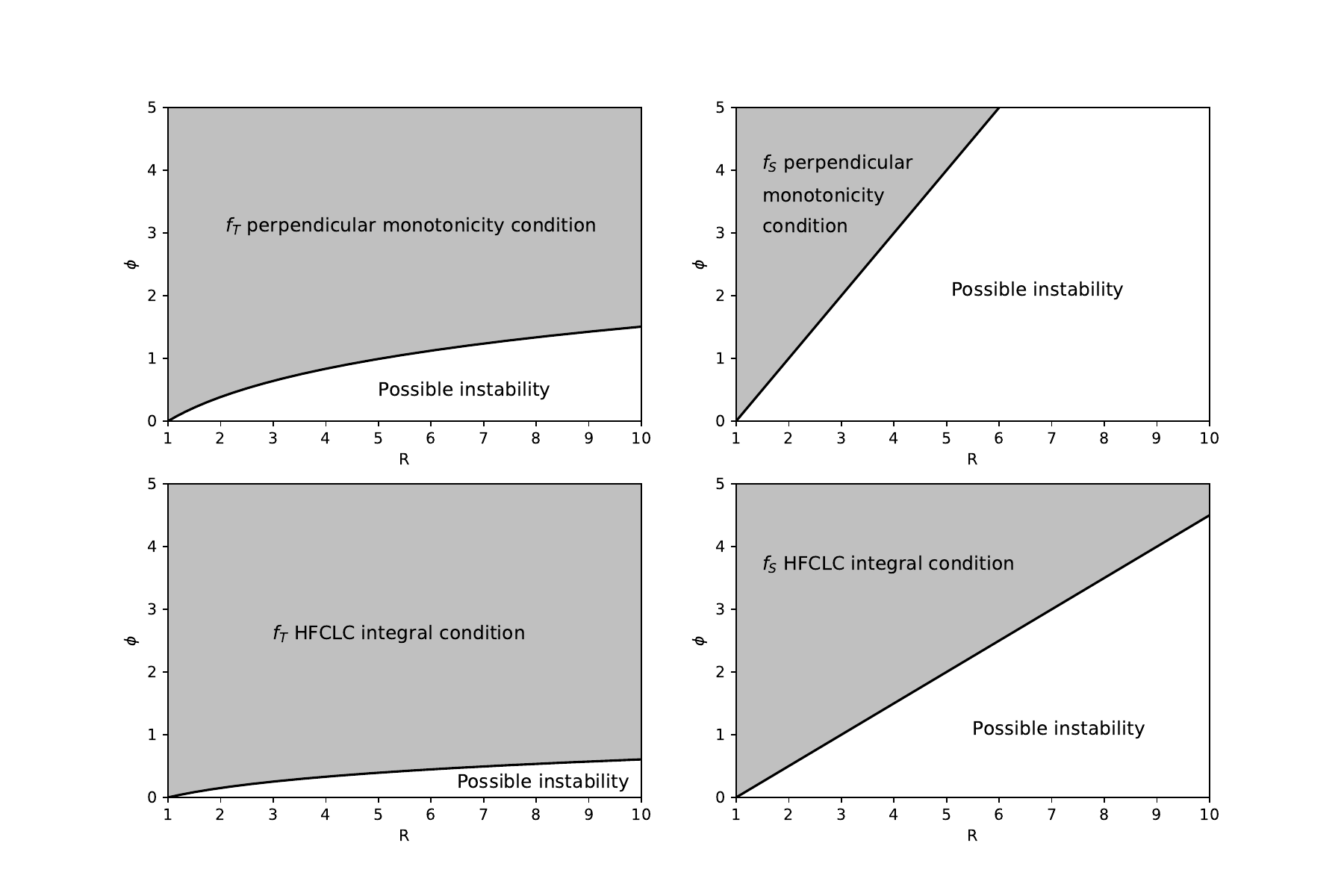}
    \caption{The first and second HFCLC stability conditions for two choices of model distribution. Either condition is sufficient for stability; the integral condition is more easily satisfied than the monotonicity condition.}
    \label{fig:HFCLC_stability}
\end{figure}

The second sufficient condition is somewhat more involved to evaluate. 
It is convenient to change integration variables to $u \doteq \sqrt{x-y^2}$, so that 
\begin{gather}
\int_{y^2}^\infty \big( x - y^2 \big)^{-1/2} \frac{\partial \psi}{\partial x} \, \D x = 2 \int_0^\infty \frac{\partial \psi}{\partial x} \, \D u .
\end{gather}
When $\phi \geq 0$, the second sufficient condition for $\psi_T$ can be written as 
\begin{align}
&\int_0^\infty e^{-u^2-y^2} \text{erf} \big[ \sqrt{(R-1) (u^2+y^2) + \phi} \, \big] \D u \nonumber \\
&\quad \geq \int_0^\infty \frac{(R-1) e^{-R(u^2+y^2)-\phi}}{\sqrt{\pi} \sqrt{(R-1)(u^2+y^2)+\phi}} \, \D u \quad \forall y > 0. 
\end{align}
The RHS integral can be evaluated, so that this is equivalent to 
\begin{align}
&\int_0^\infty e^{-u^2-y^2} \text{erf} \big[ \sqrt{(R-1)(u^2+y^2)+\phi} \, \big] \D u \\
&\geq \frac{\sqrt{R-1}}{2\sqrt{\pi}} \exp \bigg[ \frac{1}{2} \bigg( -R y^2 - 2 \phi + \frac{R \phi}{R-1} \bigg) \bigg] K_0 \bigg[ \frac{R}{2} \bigg( y^2 + \frac{\phi}{R-1} \bigg) \bigg] \quad \forall y > 0. 
\end{align}
Here $K_0$ is the modified Bessel function of the second kind. 
This stability condition is calculated numerically in Figure~\ref{fig:HFCLC_stability}. 

The second condition for $\psi_S$ is 
\begin{align}
&\int_0^\infty e^{-x} \bigg\{ (R-1) \big[ (R-1) x + \phi \big] \delta \big[ (R-1) x + \phi \big] \nonumber \\
&\hspace{30 pt} - \big[ (R-1)(x-1) + \phi \big] \Theta \big[ (R-1)x + \phi \big] \bigg\} \D u \leq 0 \quad \forall y > 0. 
\end{align}
When $\phi \geq 0$, the condition becomes 
\begin{gather}
\int_0^\infty e^{-u^2-y^2} \big[ (R-1) (u^2+y^2-1) + \phi \big] \D u \geq 0 \quad \forall y > 0. 
\end{gather}
This integral can be evaluated explicitly: 
\begin{gather}
\int_0^\infty e^{-u^2-y^2} \big[ (R-1) (u^2+y^2-1) + \phi \big] \D u = \frac{\sqrt{\pi}}{4} e^{-y^2} \big[ (R-1) (2 y^2 - 1) + 2 \phi \big] , 
\end{gather}
so the condition is satisfied when 
\begin{gather}
\phi \geq \frac{R-1}{2} \, . 
\end{gather}
To check the condition when $\phi < 0$, it is sufficient to evaluate the integral when $y$ is less than $\sqrt{-\phi/(R-1)}$: 
\begin{align}
&\int_{\sqrt{-\phi/(R-1)}}^\infty e^{-u^2-y^2} \big[ (R-1) (u^2+y^2-1)+\phi \big] \D u \nonumber \\
&\hspace{50 pt} = e^{-y^2} \bigg\{ \frac{1}{2 \sqrt{2}} e^{\phi/(R-1)} \sqrt{-2 \phi(R-1)} \nonumber \\
&\hspace{130 pt}- \frac{\sqrt{\pi}}{4} \big[ (R-1)(1-2y^2)-2\phi\big] \text{erfc} \bigg[ \sqrt{\frac{-\phi}{R-1}} \,
\bigg] \bigg\} . 
\end{align}
Note that for $R > 1$ and $\phi < 0$, 
\begin{gather}
\frac{1}{2 \sqrt{2}} < \frac{\sqrt{\pi}}{4} \\
\sqrt{-(R-1) 2 \phi} < R - 1 - 2 \phi \\
e^{\phi/(R-1)} < \text{erfc} \sqrt{\frac{-\phi}{R-1}} \, , 
\end{gather}
so the second sufficient condition is never met for $\psi_S$ when $\phi < 0$. 

These conditions are shown for both model distributions in Figure~\ref{fig:HFCLC_stability}. 
A higher mirror ratio $R$ consistently makes the HFCLC mode more difficult to stabilize. 
This makes sense; the population inversion appears through the projected perpendicular distribution $\psi(x)$, and higher $R$ makes the loss-cone structure steeper in that distribution. 

Especially at higher $R$, the truncated Maxwellian model $f_T$ is stabilized much more easily than the smoothed polynomial prefactor model $f_S$. 
This makes sense; as $\phi$ becomes larger, $f_S$ remains anisotropic even far from the loss-cone boundary itself, whereas $f_T$ reverts to a Maxwellian everywhere away from the boundary. 

There are a variety of additional corrections to the stability condition not considered here. 
\citet{Turikov1973} discusses a density-dependent correction to the centrifugal stabilization effect that becomes more important at lower densities. 
Nonzero electron temperature makes the HFCLC mode easier to stabilize \citep{Post1987}. 
So does any other modification of the ion distribution that helps to fill in the population inversion, such as the ``warm plasma stabilization'' concept \citep{Post1967, Post1987}. 
Finite-geometry effects may also help to stabilize the mode \citep{Rosenbluth1965, Aamodt1966, Gerver1979}. 

\subsection{Stability of Distributions from Fokker Planck Simulations}

\begin{figure}
    \centering
    \includegraphics[width=0.9\linewidth]{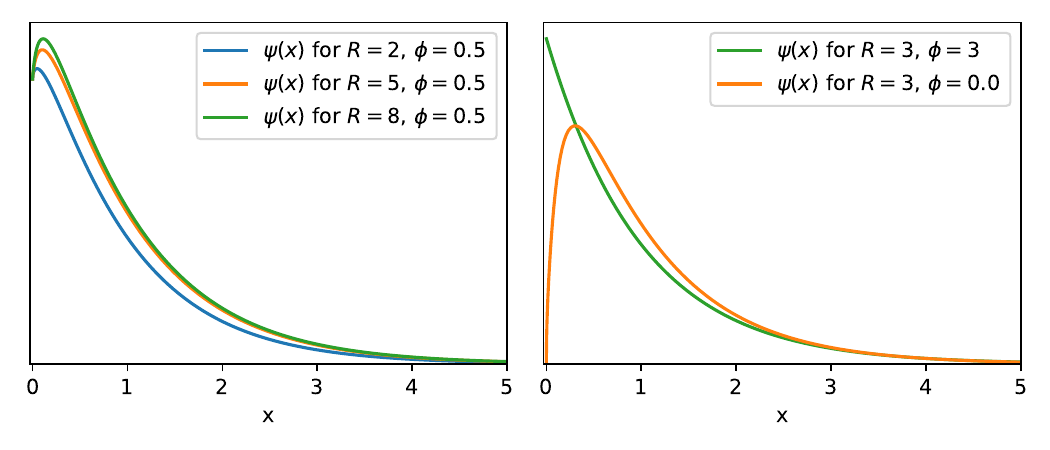}
    \caption{Projected perpendicular distribution $\psi(x)$ (unnormalized, arbitrary units) from Fokker-Planck simulations for several values of the mirror ratio and confining potential.
    Comparing to Fig.~\ref{fig:psiPlots}, we see that the behavior much more closely matches the truncated Maxwellian model $f_T$ than the Volosov model $f_S$.}
    \label{fig:fp_psi}
\end{figure}

To compare the validity of the two models, we can also generate equilibria using Fokker-Planck simulations.
We consider the thermally-normalized, sourced, steady-state diffusion equation in pitch angle $\theta$ and normalized velocity $\xv \doteq |v|/\sqrt{2 T / m}$  \citep{Pastukhov1974,Najmabadi1984,Ochs2023ConfinementTime}:
\begin{align}
	0 &=  \pa{}{\xv} \left[\sqrt{g}\left( \frac{\xv}{\xv^3+\xv_0^3} f + \frac{1}{2 (\xv^3+\xv_0^3)} \pa{f}{\xv}\right)\right] +  \pa{}{\theta} \left[\sqrt{g} \left(\frac{Z_\perp }{\xv^3+\xv_0^3}  \pa{f}{\theta} \right)\right]  + \sqrt{g} s(\xv,\theta)  ;\label{eq:diffusionEquationSim}\\
	\sqrt{g} &\doteq 4\pi \xv^2 \sin \theta; \qquad s \doteq \pi^{-3/2} e^{-\xv^2/T_s},
\end{align}
with reflecting boundary conditions everywhere except the loss cone, determined by:
\begin{align}
    R \sin^2 \theta &= 1- \frac{\phi}{\xv^2}. \label{eq:lossCone}
\end{align}
For $Z = 1$ (singly charged) ions, $Z_\perp = 1/2$, and we additionally take $\xv_0 = 0.01$.\footnote{The roles of $Z_\perp$ and $z_0$ in the collision operator are discussed in further detail in \citet{Ochs2023ConfinementTime}. $Z_\perp$ helps to control the relative rates of different collisional processes, and $z_0$ helps to prevent unphysical divergences in the collision operator. For present purposes, they are treated as fixed parameters. }
We also allow for different source temperatures $T_s$ relative to the normalized temperature towards which collisions drive the distribution function.

This collision model is a simplified form of the asymptotic high-velocity limit of the Rosenbluth potentials.
Although not the most accurate model for thermal ion-ion collisions, it has the nice feature that it very rapidly drives the low-energy part of the distribution to a Maxwellian, due to the high diffusion coefficient, while preserving the diffusion-influenced structure of the solution near the loss cone.
For implementation, we use the Fokker-Planck code based on DolfinX \citep{Scroggs2022ConstructionArbitrary} and gmsh \citep{Geuzaine2009Gmsh3D} that was developed in \cite{Ochs2023ConfinementTime}.
We use a mesh size of $\Delta \xv = 0.025$.

\begin{figure}
    \centering
    \includegraphics[width=0.6\linewidth]{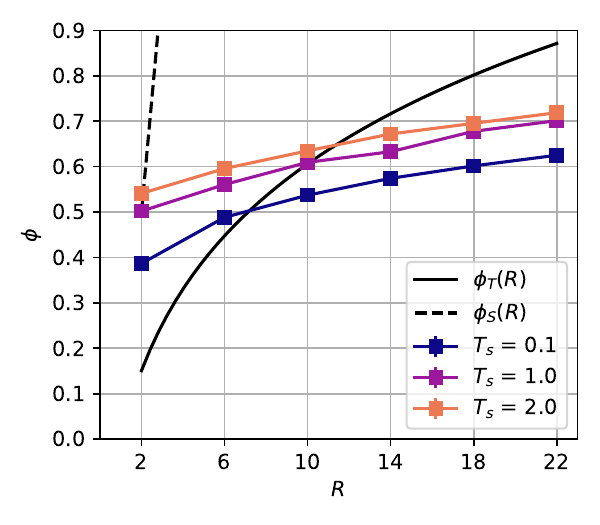}
    \caption{HFCLC stability boundary for the Fokker-Planck simulations, with sources at different temperatures, plotted alongside the truncated Maxwellian model threshold $\phi_T$ (corresponding to distribution $f_T$) and Volosov's model $\phi_S$ (corresponding to distribution $f_S$).
    Estimated error bars based on pseudo-error scaling are plotted, but fall within the square markers. 
    The pseudo-error is calculated by varying the mesh size $\Delta z$, calculating the rate of convergence of the results, and using this to infer the numerical error at the chosen resolution. 
    The qualitative trend of decreasing stability at greater $R$ remains true in the model, though with a less steep slope than for $\phi_T$.
    $\phi_S$, meanwhile, very quickly diverges from the simulation lines. 
    See Eq.~(\ref{eqn:phiMa}), and the discussion of the ambipolar potential that follows it, for more on the mapping between $\phi$ and the rotational Mach number. }
    \label{fig:fp_hfclc_stability}
\end{figure}

The stability boundary results of these simulations for the HFCLC stability criterion (Eq.~(\ref{eqn:hfclc})) are shown in Figure~\ref{fig:fp_hfclc_stability}, alongside the truncated Maxwellian and Volosov models, for several values of the normalized source temperature $T_s$. 
The simulations show the same general trend of decreasing stability with greater $R$ as the truncated Maxwellian model, but the $R$ dependence in the Fokker-Planck simulations is noticeably weaker. 
To some extent, this is to be expected. 
In the truncated Maxwellian model, the effect of the loss cone scales with the volume of the loss cone in phase space (since that region is simply empty whereas the rest of the distribution is unaffected). 
The numerical results include the effects that the loss cone has on the population outside of the loss cone itself. 
If there is any loss region at a given $v$, pitch-angle scattering into that region will have some tendency to suppress the whole distribution at $v$. 
So, although the stability boundary predicted from the truncated Maxwellian is qualitatively similar to the simulated results, they are not a perfect match. 
Note that this dependence is on $R$ for fixed $\phi$, not fixed Mach number. 
Apart from that, the simulations make it clear that a higher-temperature source somewhat destabilizes the distribution (as expected). 
Finally, it is clear that the Volosov model does not remotely match the results of the simulations at high $R$.

As before, we can also examine the projected perpendicular energy distribution $\psi(x)$, where $x = v_\perp^2/\bar v_i^2$, for the simulation results for set values of the mirror ratio and confining potential. This analysis is shown in Fig.~\ref{fig:fp_psi}.
Comparing to Fig.~\ref{fig:psiPlots}, we see that once again, the curves qualitatively match the behavior of the truncated Maxwellian model $f_T$, but not of the Volosov model $f_S$.

\section{Drift Cyclotron Loss Cone Mode} \label{sec:DCLC}

The drift cyclotron loss cone mode (DCLC) is another loss-cone instability. 
In its simplest form, it is electrostatic and has vanishing $k_{||}$ \citep{Post1966}. 
As was the case for the HFCLC, small $k_{||}$ is the result of the electron mobility. 
The DCLC, like the HFCLC, draws free energy from the loss-cone population inversion. 
Unlike the HFCLC, it also draws free energy from the presence of a (spatial) density gradient. 
Both factors must be present for the DCLC to occur as an unstable mode. 
Instability requires that the gradient scale length $|\nabla n / n|^{-1}$ not be very much larger than the ion Larmor radius \citep{Kotelnikov2017}. 
The DCLC is related to the drift cyclotron (DC) instability, which draws only on the density gradient \citep{Mikhailovsky1965}. 
Intuitively, the DCLC can be understood as a kind of drift wave that is destabilized by the presence of the loss cone. 

The short-wavelength electrostatic $k_{||}=0$ DCLC for a slab is probably best known in terms of the following dispersion relation \citep{Post1966, Post1987, Kotelnikov2017}:
\begin{gather}
w^2 \cot w + \beta w = \beta^{2/3} (\varepsilon a_i) \pi^{4/3} \bigg( \frac{Z m_e}{m_i} + \frac{\omega_{ci}^2}{\omega_{pi}^2} \bigg)^{-2/3} \label{eqn:standardDCLCDispersion}
\end{gather}
where $w \doteq \pi \omega / \omega_{ci}$, $Z$ is the ion charge state, and 
\begin{gather}
\beta \doteq \pi (k a_i)^3 \bigg( \frac{Z m_e}{m_i} + \frac{\omega_{ci}^2}{\omega_{pi}^2} \bigg) . \label{eqn:standardBeta}
\end{gather}
In Eq.~(\ref{eqn:standardDCLCDispersion}), $\varepsilon$ is the gradient scale length of the density $n$; in Cartesian $(x,y,z)$ coordinates, if $\nabla n$ is directed in the $y$ direction, it can be written as 
\begin{gather}
\varepsilon \doteq \frac{1}{n} \frac{\partial n}{\partial y} \, . 
\end{gather}
Finally, $a_i$ is typically given by 
\begin{gather}
a_i = \frac{1}{\Omega_i} \bigg[ \int \frac{f_i}{v_\perp^3} \, \D^3 \bv \bigg]^{-1/3} . \label{eqn:standardAi}
\end{gather}
As written, nothing about this dispersion relation suggests that the monotonicity of $f_i$ should play any special role in the stability of the mode. 
The shape of the distribution comes into the problem through $a_i$, and it turns out that the value of $a_i$ determines the minimal unstable gradient scale length $\varepsilon$ \citep{Post1966}: 
\begin{gather}
(\varepsilon \, a_i)_\text{crit} = 0.38 \bigg( \frac{Z m_e}{m_i} + \frac{\omega_{ci}^2}{\omega_{pi}^2} \bigg)^{2/3} . 
\end{gather}
Suppression of the loss cone does not appear to be stabilizing except in this limited sense. 

However, it turns out that Eq.~(\ref{eqn:standardDCLCDispersion}) is derived assuming that $f_i$ must vanish as $v_\perp \rightarrow 0$. 
The derivation of Eq.~(\ref{eqn:standardDCLCDispersion}) includes two steps in which boundary terms involving $f_i |_{v_\perp = 0}$ are discarded. 
Eq.~(\ref{eqn:standardDCLCDispersion}) is obtained by integrating over unperturbed particle trajectories; see, for example, the discussions in \citet{Post1966}, \citet{Swanson}, or \citet{Kotelnikov2017}. 
The dispersion relation can be written in terms of the electron and ion susceptibilities $\chi_e$ and $\chi_i$ as 
\begin{gather}
0 = 1 + \chi_e + \chi_i. 
\end{gather} 
The electron susceptibility is unchanged from what can be found in previous sources: 
\begin{gather}
\chi_e = - \frac{\omega_{pe}^2}{\omega_{ce}^2} - \frac{\omega_{pe}^2 \varepsilon}{k_\perp \omega \omega_{ce}}
\end{gather} 
where $k = k_\perp$ is the (purely perpendicular) wavenumber. 
The ion contribution comes from the following: 
\begin{align}
\chi_i &= \frac{2 \pi \omega_{pi}^2}{k^2} \int_{-\infty}^{+\infty} \D v_{||} \int_0^\infty v_\perp \, \D v_\perp \sum_{n=-\infty}^{\infty} \frac{J_n^2(k v_\perp / \omega_{ci})}{\omega + n \omega_{ci}} \frac{n \omega_{ci}}{v_\perp} \frac{\partial f_i}{\partial v_\perp} 
\end{align}
where $J_n$ is a Bessel function of the first kind. 
Using the identity that $\sum_{n=-\infty}^{\infty} J_n^2(x) = 1$, this can be rewritten as 
\begin{gather}
\chi_i = \frac{2 \pi \omega_{pi}^2}{k^2} \int_{-\infty}^{+\infty} \D v_{||} \int_0^\infty \D v_\perp \, \frac{\partial f_i}{\partial v_\perp} \bigg[ 1 - \sum_{n=-\infty}^{\infty} \frac{\omega J_n^2(k v_\perp / \omega_{ci})}{\omega + n \omega_{ci}} \bigg] . \label{eqn:chiISplitSum}
\end{gather}
Then, following \citet{Post1966}, the sum can be evaluated in the short-wavelength limit by considering the asymptotic behavior of the Bessel functions: 
\begin{align}
\sum_{n=-\infty}^{\infty} \frac{\omega J_n^2(k v_\perp / \omega_{ci})}{\omega + n \omega_{ci}} &\approx \frac{\omega}{\pi k v_\perp} \sum_{n=-\infty}^{\infty} \frac{1}{n + \omega / \omega_{ci}} \\
&= \frac{\omega}{k v_\perp} \cot \bigg( \frac{\pi \omega}{\omega_{ci}} \bigg). 
\end{align}
(Here we have taken $k$ to be positive.) 
This leads to 
\begin{align}
\chi_i &= \frac{2 \pi \omega_{pi}^2}{k^2} \int_{-\infty}^{+\infty} \D v_{||} \int_0^\infty \D v_\perp \, \frac{\partial f_i}{\partial v_\perp} \bigg[ 1 - \frac{\omega}{k v_\perp} \cot \bigg( \frac{\pi \omega}{\omega_{ci}} \bigg) \bigg] \\
&= - \frac{2 \pi \omega_{pi}^2}{k^2} \int_{-\infty}^{+\infty} \D v_{||} \bigg[ f_i \big|_{v_\perp = 0} + \frac{\omega}{k} \cot \bigg( \frac{\pi \omega}{\omega_{ci}} \bigg) \int_0^\infty \D v_\perp \, \frac{1}{v_\perp} \frac{\partial f_i}{\partial v_\perp} \bigg]. 
\end{align}
For cases in which $f_i$ vanishes when $v_\perp$ is small, the first term in square brackets vanishes and the second can be integrated by parts to get the form of $a_i$ given in Eq.~(\ref{eqn:standardAi}). 
For more general distributions, Eq.~(\ref{eqn:standardDCLCDispersion}) becomes 
\begin{gather}
w^2 \cot w + \beta w = \beta^{2/3} (\varepsilon a_i) \pi^{4/3} \bigg( \frac{Z m_e}{m_i} + \frac{\omega_{ci}^2}{\omega_{pi}^2} + \frac{2 \pi \omega_{ci}^2}{k^2} \int_{-\infty}^{+\infty} \D v_{||} \, f_i \big|_{v_\perp = 0} \bigg)^{-2/3} \label{eqn:generalDCLC}
\end{gather}
with 
\begin{gather}
\beta = \pi (k a_i)^3 \bigg( \frac{Z m_e}{m_i} + \frac{\omega_{ci}^2}{\omega_{pi}^2} + \frac{2 \pi \omega_{ci}^2}{k^2} \int_{-\infty}^{+\infty} \D v_{||} \, f_i \big|_{v_\perp = 0} \bigg) \label{eqn:generalBeta}
\end{gather}
and 
\begin{gather}
a_i = \frac{1}{\omega_{ci}} \bigg[ \int \frac{1}{v_\perp^2} \frac{\partial f_i}{\partial v_\perp} \, \D^3 \bv \bigg]^{-1/3}. \label{eqn:generalAi}
\end{gather}
This can be understood as a generalization of the original DCLC dispersion relation to include more general distributions. 
Eq.~(\ref{eqn:generalDCLC}) can also be written as 
\begin{gather}
w^2 \cot w + \beta w = \pi^2 k^2 \varepsilon a_i^3. 
\end{gather}
Eqs.~(\ref{eqn:generalBeta}) and (\ref{eqn:generalAi}) are the same coefficients $\beta$ and $a_i$ described by Eqs.~(\ref{eqn:standardBeta}) and (\ref{eqn:standardAi}), evaluated in the more general case where $f_i$ need not vanish as $v_\perp \rightarrow 0$. 
The form of $a_i$ described by Eq.~(\ref{eqn:standardAi}) is always positive, but the more general $a_i$ in Eq.~(\ref{eqn:generalAi}) can be negative, and is guaranteed to be negative any time $\int f_i \, \D v_{||}$ is a monotonically decreasing function of $v_\perp$. 
When $a_i > 0$, there is always a threshold value of $\varepsilon$ beyond which this dispersion relation yields an unstable DCLC mode. 
This is no longer guaranteed when $a_i$ is negative. 

To see this, it is helpful to go back to the original argument for the DCLC critical gradient given by \cite{Post1966}. 
This goes as follows. 
Suppose $\beta > 0$ (as is always the case when $a_i > 0$). 
Then $w^2 \cot w + \beta w$ has a local maximum between $w = 0$ and $w = \pi$. 
This implies that there is a value of $\varepsilon$ beyond which two real $w$ solutions vanish. 
The vanishing of these two real solutions corresponds to the appearance of two complex solutions (one of which will have an imaginary part of each sign), and the onset of instability. 

Such a critical gradient can always be found so long as $\beta > 0$, and $\beta > 0$ so long as $a_i > 0$. 
However, for more general $\beta$, the local maximum persists only if $\beta > -1$. 
For $\beta < -1$, there is no critical gradient at which the number of unstable modes changes, and we can consider the DCLC stabilized. 
This constitutes a sufficient condition for stability, which can be written as follows: 

\textit{DCLC Integral Condition:} 
\begin{gather}
\frac{\pi k^3}{\omega_{ci}^3} \bigg[ \int \frac{1}{v_\perp^2} \frac{\partial f_i}{\partial v_\perp} \, \D^3 \bv \bigg]^{-1} \bigg( \frac{Z m_e}{m_i} + \frac{\omega_{ci}^2}{\omega_{pi}^2} + \frac{2 \pi \omega_{ci}^2}{k^2} \int_{-\infty}^{+\infty} \D v_{||} f_i \big|_{v_\perp = 0} \bigg) \leq -1 . 
\end{gather}
Recall that $k > 0$ by assumption. 
The condition can be rewritten as 
\begin{gather}
0 \leq - \int \frac{1}{v_\perp^2} \frac{\partial f_i}{\partial v_\perp} \, \D^3 \bv \leq \frac{\pi k^3}{\omega_{ci}^3} \bigg( \frac{Z m_e}{m_i} + \frac{\omega_{ci}^2}{\omega_{pi}^2} + \frac{2 \pi \omega_{ci}^2}{k^2} \int_{-\infty}^{+\infty} \D v_{||} f_i \big|_{v_\perp = 0} \bigg) . 
\end{gather}
As we saw for the HFCLC, it is the population inversion in the projection $\int f \, \D v_{||}$ that matters. 
In terms of the perpendicular energy distribution $\psi$ defined by Eq.~(\ref{eqn:psiDefinition}), this can be rewritten as 
\begin{gather}
0 \geq \int_0^\infty x^{-1/2} \frac{\partial \psi}{\partial x} \, \D x \geq - \frac{\pi k^3 \bar v_i^3}{2 \omega_{ci}^3} \bigg( \frac{Z m_e}{m_i} + \frac{\omega_{ci}^2}{\omega_{pi}^2} + \frac{2 \omega_{ci}^2}{k^2 \bar v_i^2} \, \psi(0) \bigg) . \label{eqn:DCLCIntegralConditionPsi}
\end{gather}
The first inequality in (\ref{eqn:DCLCIntegralConditionPsi}) is the same as the HFCLC integral condition when $y \rightarrow 0$. 
For the HFCLC integral condition, the choice of $y$ adjusts which parts of $\partial \psi / \partial x$ are weighted most heavily. 
For the model distributions $f_S$ and $f_T$, positive $\partial \psi / \partial x$ (when it happens) is concentrated at the smallest values of $x$. 
Therefore, the HFCLC condition and the first inequality in the DCLC condition yield the same results for $f_S$ and $f_T$. 
However, one could construct a distribution for which this would not be the case (for example, a distribution for which $\partial \psi / \partial x$ went from negative to positive to negative again as $x$ increased). 

The second inequality in the DCLC integral condition is less intuitive. 
However, it may be a less serious limitation than it initially appears. 
Note that the DCLC integral condition is satisfied any time 
\begin{gather}
0 \leq - \int_{-\infty}^{+\infty} \D v_{||} \int_0^\infty \frac{\omega_{ci}}{k v_\perp} \frac{\partial f_i}{\partial v_\perp} \, \D v_\perp \leq - \pi \int_{-\infty}^{+\infty} \D v_{||} \int_0^\infty \frac{\partial f_i}{\partial v_\perp} \, \D v_\perp . \label{eqn:DCLCIntegralConditionStronger}
\end{gather}
Recall that the dispersion relation used in this section was derived using a short-wavelength assumption. 
Specifically, in Eq.~(\ref{eqn:chiISplitSum}), the integrand was simplified by taking the asymptotic limit $k v_\perp / \omega_{ci} \gg 1$. 
The two integrands in (\ref{eqn:DCLCIntegralConditionStronger}) differ by this same factor. 

This tells us two things. 
First, within the limits of applicability of the dispersion relation, the same perpendicular monotonicity condition discussed for the HFCLC guarantees that the DCLC integral condition will also be satisfied. 
Second, even in cases where $\psi$ may not be monotonic, the second inequality in the DCLC integral condition is not a very strong constraint. 
A distribution can satisfy the first inequality in (\ref{eqn:DCLCIntegralConditionPsi}) but fail to satisfy the second only if $k v_\perp / \omega_{ci}$ is not too large for $v_\perp$ on the scale on which $f_i$ varies; it is not clear that the underlying dispersion relation will still be valid in such a case. 

Numerically, it turns out that the stability boundaries from the HFCLC integral condition and the DCLC integral condition coincide for the cases shown in Figures~\ref{fig:fp_psi} and \ref{fig:fp_hfclc_stability}, for the same reason that they coincide for the analytic models $f_T$ and $f_S$: the regions in which $\psi$ has the most positive slope are concentrated at low $v_\perp$. 
Therefore, Figure~\ref{fig:fp_hfclc_stability} can also be read as the stability boundary for the DCLC integral condition. 

As was the case for the HFCLC analysis, there are a variety of additional effects not considered here, some of which can be stabilizing for practical devices. 
These include finite-$\beta$ effects, finite system size, and the same ``warm plasma stabilization'' mentioned in Section~\ref{sec:HFCLC} \citep{Tang1972, Gerver1976, Berk1976, Kotelnikov2017}. 
We have also not considered the effects of multiple-ion-species distributions, which can be nontrivial \citep{Kotelnikov2018}. 

\section{Dory-Guest-Harris Mode} \label{sec:DGH}

The last mode we will discuss in detail is the Dory-Guest-Harris (or DGH) mode \citep{Dory1965}. 
In particular, we will focus on the electrostatic $k_{||}=0$ ``zero-frequency'' DGH mode, in which $\text{Re}(\omega)$ vanishes. 
This is not the most general form of the mode; see, for example, \citet{Callen1971}. 
However, it has the advantage of simplicity, and it was the focus of much of the original work on the subject. 
Physically, this mode can be understood in terms of single-particle drifts in an inhomogeneous electric field \citep{Post1987}. 
An appropriate wave field can cause corrections to the homogeneous $\bE \times \bB$ drift that cause particles with different charges to move in opposite directions. 
Depending on the kinetic distribution of particles, this can amplify these waves and lead to instability. 
As was the case for the HFCLC and DCLC, loss-cone distributions can have the necessary characteristics to drive the instability. 

If the mode is driven by an ion loss cone, the ion contribution follows from Eq.~(\ref{eqn:chiISplitSum}). Then the relevant dispersion relation is 
\begin{gather}
1 + \frac{\omega_{pe}^2}{\Omega_e^2} = \frac{2 \pi \omega_{pi}^2}{k^2} \int_{-\infty}^{+\infty} \D v_{||} \int_0^\infty \D v_\perp \, \frac{\partial f_i}{\partial v_\perp} \bigg[ 1 - \sum_{n=-\infty}^{\infty} \frac{\omega J_n^2(k v_\perp / \omega_{ci})}{\omega + n \omega_{ci}} \bigg] .
\end{gather}
A marginally stable distribution with $\text{Re}(\omega) = 0$ satisfies 
\begin{gather}
1 + \frac{\omega_{pe}^2}{\Omega_e^2} = \frac{2 \pi \omega_{pi}^2}{k^2} \int_{-\infty}^{+\infty} \D v_{||} \int_0^\infty \D v_\perp \, \frac{\partial f_i}{\partial v_\perp} \bigg[ 1 - J_0^2(k v_\perp / \omega_{ci}) \bigg] . \label{eqn:DGHDispersion}
\end{gather}
An essentially equivalent expression was found by \citet{Post1966}. 
If the RHS of Eq.~(\ref{eqn:DGHDispersion}) is positive, then the zero-frequency DGH mode is stable \citep{Post1966}. 
Note that $J_0^2(x) \leq 1$ for all $x$. 
It follows immediately that the same monotonicity condition we have discussed for the HFCLC and DCLC modes -- that is, Eq.~(\ref{eqn:monotonicityCondition}) -- is a sufficient condition for the stability of this mode as well. 
A more easily met sufficient condition for stability is to have 
\begin{gather}
\int_{-\infty}^{+\infty} \D v_{||} \int_0^\infty \D v_\perp \, \frac{\partial f_i}{\partial v_\perp} \big[ 1 - J_0^2(k v_\perp / \omega_{ci}) \big] \leq 0. 
\end{gather}
In terms of the perpendicular energy distribution $\psi$, this stability condition can be written as follows. 

\textit{DGH Integral Condition:}
\begin{gather}
\int_0^\infty \frac{\partial \psi}{\partial x} \big[ 1 - J_0^2 (\xi x^{1/2}) \big] \D x \leq 0 \label{eqn:DGHCondition}
\end{gather}
where 
\begin{gather}
\xi \doteq \frac{k \bar v_i}{\omega_{ci}} \, . 
\end{gather}
The same stability condition can be written equivalently as 
\begin{gather}
\int_0^\infty \xi x^{-1/2} J_0 (\xi x^{1/2}) J_1(\xi x^{1/2}) \psi(x) \, \D x \geq 0. 
\end{gather}
This condition was also discussed by \citet{Post1966}. 

The DGH mode is much easier to stabilize than the HFCLC or DCLC modes. 
Consider the implications of the Bessel function weighting factor in Eq.~(\ref{eqn:DGHCondition}). 
One minus the square of the Bessel function $J_0^2$ vanishes when its argument is small, then exhibits damping oscillations with a mean that tends toward unity as the argument increases. 
Mirror-like equilibria, including both $f_T$ and $f_S$, typically have their most destabilizing parts (that is, the regions of velocity space in which the population inversion is steepest) at small $v_\perp$. 
This can be seen in Figure~\ref{fig:psiPlots}. 
The exception, for $f_S$ and $f_T$, is when $\phi$ is negative (when the potential is deconfining rather than confining). 
When $\phi < 0$, the regions with the most positive $\partial \psi / \partial x$ are shifted to larger values of $x$, below which the distribution entirely vanishes, and it is possible to find choice of $R$ and $\xi$ for which the DGH integral condition is not satisfied. 
However, $\phi < 0$ is not the case that is relevant for centrifugal confinement. 

The numerical solutions described in Section~\ref{sec:HFCLC} are also stable to the DGH in most cases. 
However, it is possible to find DGH-unstable equilibria in cases with relatively high-temperature sources (for example, $T_s = 2$) and little rotation ($\phi < 1/2$). 
Because the numerical instability threshold is strongly determined by the source term, and because the DGH is consistently stabilized at a $\phi$ threshold below that shown in Figure~\ref{fig:fp_hfclc_stability}, we will not focus on this effect here. 
The two main things that are worth noting, for present purposes, are (1) that the DGH is stable for $\phi \geq 0$ for the analytic models but not always for the numerical simulations and (2) that even in the numerical simulations the DGH still appears to be more easily stabilized than either the HFCLC or the DCLC. 

\section{Free Energy and Stability of Flute-Like Modes} \label{sec:fluteRearrangements}

A common thread across all of these modes is that stability is guaranteed whenever $\psi$ is a monotonically decreasing function: that is, whenever 
\begin{gather}
\frac{\partial}{\partial v_\perp^2} \int_{-\infty}^{+\infty} f(v_\perp^2, v_{||}^2) \, \D v_{||} \leq 0. \label{eqn:projectedMonotonicity}
\end{gather}
As we have seen, this condition is sufficient but not necessary for each of the modes in question. Nonetheless, it is interesting because of its simplicity and because of the fact that it appears in the stability analyses for a range of different modes. 

Another commonality (which applies to all loss-cone modes, not just the ones considered here) is that the physical intuition for how they work is typically posed in terms of free energy and phase-space rearrangements. 
Systems with loss cones typically have population inversions. 
This means that energy can be released by rearranging the contents of phase space such that high-energy particles move to lower energy. 
This energy then becomes available to drive instabilities. 

This intuition is closely related to the idea underlying the theory of plasma free energy, wherein the capacity of a system for instability is set by the amount of energy that can be released by certain classes of phase-space rearrangements. 
Given a set of rules for which phase-space rearrangements are allowed, it is possible to calculate the free energy (sometimes called the ``available energy'') by determining the maximal energy that can be extracted from the system using these rearrangements. 
This energy could then drive instabilities, or one could imagine extracting it intentionally, as in the case of alpha-channeling \citep{Fisch1992}. 

Several different classes of rearrangement have been proposed, each corresponding to a different way of defining the free energy. 
\citet{Gardner1963} suggested a formulation that is now sometimes called ``Gardner restacking,'' in which any operation that preserves phase space densities is permitted \citep{Dodin2005, Helander2017ii, Helander2020}. 
An alternative proposed by \citet{Fisch1993}, sometimes called the ``diffusively accessible free energy,'' instead allows operations that mix or average the populations between phase space elements \citep{Hay2015, Hay2017, Kolmes2020ConstrainedDiffusion, Kolmes2020Gardner, Kolmes2022Plateau}. 
There are variants of both formulations that allow for additional restrictions to be imposed on the allowed rearrangements, enforcing a conservation law  -- for example, on an adiabatic invariant $\mu$ -- by requiring that exchange or mixing operations must act only between elements of phase space with the same value of $\mu$ \citep{Helander2017ii, Helander2020, Kolmes2020ConstrainedDiffusion}. 
Recent evidence suggests that adding suitable adiabatic invariants to the Gardner free energy leads to a metric that can predict the saturation levels of some types of turbulence \citep{Mackenbach2022, Mackenbach2023Available, Mackenbach2023Miller}. 

When we try to understand the behavior of flute-like loss-cone modes in terms of free energy theory, we arrive at an apparent paradox. 
On the one hand, free energy/rearrangement theories appear to match our intuitions for the physics behind loss-cone instabilities. 
Loss-cone instabilities happen because it is possible to release energy by rearranging the distribution in velocity space; this is exactly the kind of situation that free energy theories should be able to describe. 
On the other hand, the mirror ratio dependence found in Sections~\ref{sec:HFCLC} and \ref{sec:DCLC}, and shown in Figure~\ref{fig:HFCLC_stability}, is not at all consistent with what would have been predicted from either the Gardner or the diffusive exchange theories. 
A larger mirror ratio means a smaller loss cone, with a correspondingly smaller energy release if that loss cone is filled in, but the stabilization thresholds for the HFCLC and DCLC suggest that configurations with larger mirror ratios are actually more difficult to stabilize. 

To resolve this contradiction, it is helpful to begin by considering the condition for a distribution to be a ``ground state'' in different versions of free energy theory. 
In the absence of any additional conservation laws, a system is in a ground state with vanishing free energy only when the distribution $f$ is a monotonically decreasing function of energy. That is, if $f = f(v^2)$, and if the energy is $m v^2 / 2$, being in a ground state requires that 
\begin{gather}
\frac{\D f(v^2)}{\D v^2} \leq 0 \quad \forall v. 
\end{gather}
This holds for both the Gardner and the diffusive theories. 
If $\mu$ invariance is enforced, a distribution of the form $f = f(v^2, \mu)$ is a ground state only if \citep{Helander2017ii} \footnote{The more general ground-state condition is described in \cite{Helander2017ii}.}
\begin{gather}
\frac{\partial f(v^2, \mu)}{\partial v^2} \leq 0 \quad \forall v. 
\end{gather}
Considering the resemblance of these conditions to Eq.~(\ref{eqn:projectedMonotonicity}), and the similarity of the intuitions behind the free-energy theories and the loss-cone modes, it is natural to wonder whether the behavior of these loss-cone instabilities can be understood in terms of the free energy. 

To motivate a free energy with ground states corresponding to Eq.~(\ref{eqn:projectedMonotonicity}), consider the standard quasilinear diffusion equation describing the velocity-space diffusion induced by wave-particle interactions: 
\begin{gather}
\frac{\partial f}{\partial t} = \frac{\partial}{\partial \bv} \cdot \mathsf{D} \cdot \frac{\partial}{\partial \bv} \, f, 
\end{gather}
where $\mathsf{D}$ is a rank-2 tensor given by 
\begin{gather}
\mathsf{D} \doteq D_0 \int \frac{\omega_i \mathcal{E}_{\mathbf{k}}}{(\mathbf{k} \cdot \bv - \omega_r)^2 + \omega_i^2} \frac{\mathbf{k} \mathbf{k}}{k^2} \, \D \mathbf{k}, 
\end{gather}
for some species-dependent constant $D_0$ and spectral wave energy density $\mathcal{E}_\mathbf{k}$, and with $\omega_r$ and $\omega_i$ denoting the real and imaginary parts of the wave frequency $\omega$ \citep{KrallTrivelpiece}. 
In the limit where $k_{||}$ is vanishingly small, this diffusion operator acts only in perpendicular directions, and does not distinguish between different values of $v_{||}$. 

If we expect phase space to be rearranged by interactions with small-$k_{||}$ modes, then it is sensible to consider the free energy associated with mixing operations that can act only on the projected distribution function $\int_{-\infty}^{+\infty} f \, \D v_{||}$. 
Then the ground-state condition immediately recovers Eq.~(\ref{eqn:projectedMonotonicity}). 

\begin{figure}
    \centering
    \vspace{0.4cm}
    \includegraphics[width=\linewidth]{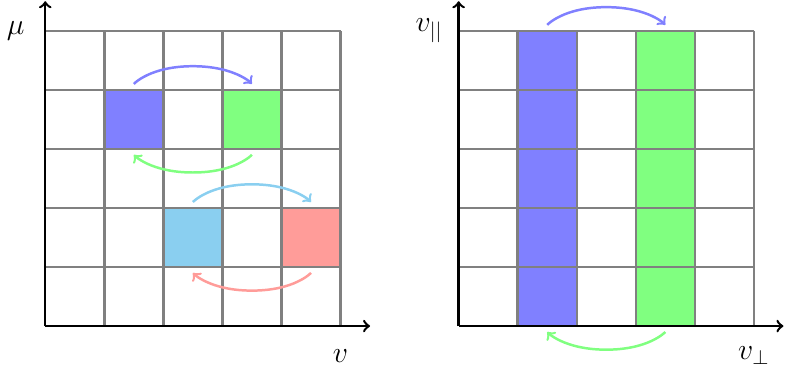}
    \caption{Two different kinds of restriction on the allowed rearrangement operations, each enforcing a conservation law. On the left, $\mu$ conservation is enforced by allowing any rearrangement that acts on two elements of phase space with the same value of $\mu$. On the right, $v_{||}$ is conserved by instead enforcing that rearrangement operations must act to exchange populations between the entire region of phase space with one value of $v_\perp$ and the entire region with another $v_\perp$.}
    \label{fig:projectionSectionCartoon}
\end{figure}

This rule bears some similarity to the restacking conservation laws introduced by Helander \citep{Helander2017ii, Helander2020}. 
Indeed, it leads to the conservation of $v_{||}$, but it is more restrictive than Helander's rule. 
The difference comes from the assumption that these flute-like modes not only cannot move rearrange particles in the parallel velocity-space direction, but they also cannot perform rearrangements that distinguish between populations with the same $v_\perp$ but different $v_{||}$. 
This distinction is illustrated in Figure~\ref{fig:projectionSectionCartoon}. 

\section{Rearrangements in Loss-Cone Systems} \label{sec:lossConeRearrangements}

Thus far, the discussion of the allowed rearrangements has not taken into account the existence of the loss cone. 
The preceding sections have considered loss-cone instabilities as those instabilities which result from the particle distributions that are characteristic of systems with loss cones, without considering the effects of the phase-space loss region itself. 
In one sense, there is nothing wrong with this; there is no reason why loss-cone-like distributions, and their associated instabilities, cannot exist without the presence of loss regions in phase space. 
(In other words, it would be self-consistent to imagine a mirror-like distribution of particles that happens to exist in an infinite homogeneous space, and to calculate its stability properties). 
However, if we want to apply free energy theory to cases in which there are regions of phase space from which particles are promptly lost, then we need to modify our accessible states accordingly. 

Once a particle enters the loss cone, it exits the system. 
This means that it should not be possible to ``fill up'' a loss cone. 
It is possible to model the energetic fate of these vanishing particles in more than one way, but the simplest approach is to assert that they leave the system without any further change in energy. 
Then a particle starting at energy $\varepsilon_i$ and leaving the system with energy $\varepsilon_f$ leaves behind energy $\varepsilon_i - \varepsilon_f$ within the system. 
This energy difference is thus ``available'' in the same sense that is usually measured by the Gardner free energy. 
The situation would be very different if there were some other pathway by which the liberated energy could exit the system. 

When describing rearrangement processes in the presence of a loss cone, it is notationally convenient (and physically equivalent) to instead say that any region of phase space within the loss cone interacts with other parts of phase space as if it had zero population. 
As an illustrative example, consider the four-state discrete system in which the states have dimensionless energies $\varepsilon_0 = 0$, $\varepsilon_1 = 1$, $\varepsilon_2 = 2$, and $\varepsilon_3 = 3$. Suppose the second-lowest-energy state is part of a region of phase space from which particles are promptly lost; we will denote this by shading it light blue. 
Then the state
\begin{gather*}
\begin{array}{|c|c|c|c|}
\hline 
f_0 & \cellcolor{blue!15} f_1 & f_2 & f_3 \\ \hline
\end{array}
\end{gather*}
has energy $\sum_i \varepsilon_i f_i$. 
If not for the presence of the loss cone, Gardner restacking would lead to a ground state simply by sorting the $f_i$ from largest to smallest. 
For example, it would take 
\begin{align*}
\begin{array}{|c|c|c|c|}
\hline 
4 & 0 & 2 & 1 \\ \hline
\end{array}
&\rightarrow 
\begin{array}{|c|c|c|c|}
\hline 
4 & 2 & 0 & 1 \\ \hline
\end{array}
\rightarrow 
\begin{array}{|c|c|c|c|}
\hline 
4 & 2 & 1 & 0 \\ \hline
\end{array}
\end{align*}
for a total energy release of $(2 \varepsilon_2 + \varepsilon_3) - (2 \varepsilon_1 + \varepsilon_2) = 3$. 
On the other hand, if the second cell is part of a loss cone, the mapping would instead be 
\begin{align*}
\begin{array}{|c|c|c|c|}
\hline 
4 & \cellcolor{blue!15} 0 & 2 & 1 \\ \hline
\end{array}
&\rightarrow 
\begin{array}{|c|c|c|c|}
\hline 
4 & \cellcolor{blue!15} 2 & 0 & 1 \\ \hline
\end{array}
\rightarrow 
\begin{array}{|c|c|c|c|}
\hline 
4 & \cellcolor{blue!15} 3 & 0 & 0 \\ \hline
\end{array}
\end{align*}
for a release of $(2 \varepsilon_2 + \varepsilon_3) - 3 \varepsilon_1 = 4$. 
The loss-cone region never runs out of space; we still imagine that phase-space densities are being conserved, because this notation is shorthand for the idea that there are many copies of that box sitting at the same energy (corresponding to a spatial coordinate as particles exit the confinement device). 
In other words, the notation describes a projection of phase space in which rearrangements into loss cones can appear to produce condensate-like accumulations of particles, even though phase space volumes are still conserved. 

Now consider the corresponding continuous problem for a simple loss-cone with no added trapping or detrapping potential. 
This is the case shown in the upper-left quadrant of Figure~\ref{fig:cartoonLossCones}. 
In this scenario, one can see that the Gardner free energy is equal to the entire kinetic energy of the system, since any populated region of phase space can exit the system through the part of the loss cone with zero energy. 
In the more general case where the lowest-energy region of the loss cone has energy $\varepsilon_\text{min} > 0$ (for example, the upper-right quadrant of Figure~\ref{fig:cartoonLossCones}), any particle that starts with energy $\varepsilon > \varepsilon_\text{min}$ can be removed from the system at $\varepsilon_\text{min}$, though the details of the ground state will also depend on the initial conditions for the parts of phase space with $\varepsilon < \varepsilon_\text{min}$. 
The situation is similar if we consider diffusive rearrangements rather than Gardner restacking; an element of phase space with initial energy above the lowest-energy region of the loss cone may be repeatedly averaged against that empty region until it has entirely been removed from the system at the minimal loss-cone energy. 

This is straightforward enough so far. 
Things become more complicated if we consider the flute-like rearrangements discussed in Section~\ref{sec:fluteRearrangements}. 
Consider, for example, a discrete system in which we distinguish between the parallel and perpendicular energy in different boxes as follows: 
\begin{gather*}
\begin{array}{|c|c|}
\hline
\cellcolor{blue!15} (\varepsilon_\perp, \varepsilon_{||}) = (0,1) & (\varepsilon_{||}, \varepsilon_\perp) = (1,1) \\
\hline
(\varepsilon_\perp, \varepsilon_{||}) = (0,0) & (\varepsilon_{||}, \varepsilon_\perp) = (1,0) \\ \hline
\end{array}
\end{gather*}
Here the state with $(\varepsilon_{||}, \varepsilon_\perp) = (1,0)$ is taken to be a loss region. 
This is roughly the kind of loss-cone structure we would expect for the loss cone of a system with $R > 1$ and some confining potential. 

Suppose we insist, following the discussion in Section~\ref{sec:fluteRearrangements}, that rearrangements must act between whole columns of states with given values of $\varepsilon_\perp$. One example of the Gardner restacking procedure might be 
\begin{gather*}
\begin{array}{|c|c|}
\hline
\cellcolor{blue!15} 0 & 3 \\
\hline
1 & 2 \\ \hline
\end{array} \rightarrow
\begin{array}{|c|c|}
\hline
\cellcolor{blue!15} 3 & 0 \\
\hline
2 & 1 \\ \hline
\end{array} . 
\end{gather*}
This produces a ground state subject to the given constraints. 
But then consider the following initial condition: 
\begin{gather*}
\begin{array}{|c|c|}
\hline
\cellcolor{blue!15} 0 & 1 \\
\hline
2 & 0 \\ \hline
\end{array} . 
\end{gather*}
This system has only one allowed exchange operation, and it leads to a higher-energy state: 
\begin{gather*}
\begin{array}{|c|c|}
\hline
\cellcolor{blue!15} 0 & 1 \\
\hline
2 & 0 \\ \hline
\end{array} \rightarrow 
\begin{array}{|c|c|}
\hline
\cellcolor{blue!15} 1 & 0 \\
\hline
0 & 2 \\ \hline
\end{array}. 
\end{gather*}
This transforms the system from a state with energy 2 to a state with energy 3. 
However, consider the following sequence: 
\begin{gather*}
\begin{array}{|c|c|}
\hline
\cellcolor{blue!15} 0 & 1 \\
\hline
2 & 0 \\ \hline
\end{array} \rightarrow 
\begin{array}{|c|c|}
\hline
\cellcolor{blue!15} 1 & 0 \\
\hline
0 & 2 \\ \hline
\end{array} \rightarrow 
\begin{array}{|c|c|}
\hline
\cellcolor{blue!15} 1 & 0 \\ \hline 
2 & 0 \\ \hline
\end{array}
\end{gather*}
This sends the system from energy 2 to energy 3, and then to energy 1. 
In other words, this initial condition cannot be transformed into any state with lower energy with any \textit{single} allowed operation, but because of the interaction between the flute-like constraint and the loss-cone, it is possible to map it to a lower-energy state by using a sequence of \textit{two} operations. 
The intermediate step, in which the energy of the system is raised, is sometimes called an ``annealing operation'' \citep{Hay2015, Kolmes2022Plateau}. 
A very similar scenario can be found if diffusive exchange operations are used in place of Gardner restacking. 

Annealing operations are never necessary to reach the ground state in the original versions of the restacking or diffusive exchange theories (that is, without this interaction between the flute-like-mode constraint and the loss cone) \citep{Hay2015}. 
Their appearance in this version of the theory suggests that we should make a distinction between two different kinds of ground states. 
The first, which we might call a weak ground state, is any state which cannot be mapped to a lower-energy state using any single allowed operation. 
The second, which we might call a strong ground state, is any state which cannot be mapped to a lower-energy state using any sequence of allowed operations. 
Of course, every strong ground state is also a weak ground state. 
The perpendicular monotonicity condition found in the context of the HFCLC, DCLC, and DGH modes corresponds to the weak ground states of the constrained system. 

\section{Conclusion} \label{sec:conclusion}

This paper consists of two interrelated parts. 
The first is a series of linear stability analyses for three electrostatic flute-like loss-cone modes: the HFCLC, DCLC, and DGH. 
For each mode, we have calculated the threshold at which rotation is expected to eliminate the instability. 
Here, the effects of rotation have been modeled as a modification of the distribution function, ``lifting'' the loss cone as centrifugal confinement enlarges the trapping region in phase space. 
The model applies equally well to any other confining potential, such as an electrostatic potential.\footnote{It would have to be modified in the case of a velocity-dependent effective potential, such as can result from RF waves. See, for example, \cite{Gormezano1979, Dodin2004, Rubin2023, Miller2023}. } 
Indeed, efforts to change the behavior of these modes (particularly the DCLC) by changing the electrostatic potential are also important. 
This is part of the reason why ``sloshing'' fast ions can be used to stabilize mirrors \citep{Kesner1973, Simonen1983, Post1987}; this approach is currently being undertaken in the Wisconsin mirror program \citep{FowlerReport}. 
We do not account for the effects of any flow shear in the system, or for any other inertial effects that can become important when the flow becomes very fast (such as Coriolis forces; see, for example, \citet{Thyagaraja2009}). 

The precise stabilization threshold varies depending on the model used for the distribution function. 
The truncated Maxwellian $f_T$ predicts that roughly sonic rotation is sufficient to stabilize all of the modes considered here, with substantially subsonic rotation being sufficient when the mirror ratio is relatively small. 
A model with a smooth polynomial cutoff (denoted by $f_S$), which has been used in previous studies, suggests that supersonic rotation becomes necessary much more quickly with increasing $R$. 
However, there are reasons to believe that this latter model is inaccurate in the limit of fast rotation, as it does not converge to a Maxwellian when the confinement becomes infinitely good. 
Numerical results using equilibria from a Fokker-Planck code are mostly in line with the predictions from the truncated Maxwellian. 
These results suggest that any device which relies on a confining potential to trap particles in the direction parallel to the magnetic field -- for example, a centrifugal mirror -- should be stable against the HFCLC, DCLC, and DGH instabilities, since the threshold for effective particle trapping is typically higher than the threshold for stabilization. 

All models agree that HFCLC and DCLC stabilization gets more difficult as the mirror ratio increases. 
This is a surprise. 
Indeed, though this trend for the HFCLC was also noted by \citet{Turikov1973}, Turikov posited that it must be an unphysical artifact of the analytic model used for the distribution.\footnote{This was specifically for Turikov's analytic calculation; accompanying numerical work in the same paper \citep{Turikov1973} showed a different $R$ dependence.}  
Na\"ively, one would have expected loss-cone modes to be most unstable when $R$ is \textit{small}; they draw their free energy from the filling-in of the loss cone, and smaller $R$ translates to a larger loss cone. 
This intuition can be expressed more formally in the language of free or available energy, which is the subject of the second part of this paper. 
In the absence of any additional constraints, both the Gardner free energy and the diffusively accessible free energy are larger when $R$ is smaller. 

This contradiction can be resolved by considering the implications of the fact that the HFCLC, DCLC, and DGH modes all have vanishing $k_{||}$ (they are flute-like). 
Quasilinear theory predicts that such modes can only rearrange velocity space in the perpendicular direction. 
Effectively, they act on the projection $\int f(v_{||}, v_\perp) \, \D v_{||}$. 
If free-energy theories are instead applied to this projection, one recovers the behavior (including the $R$ dependence) found in the linear stability analyses. 
This can be understood by noting that the projected distributions become monotonic at lower rotation speeds when $R$ is smaller, because distributions with smaller $R$ have broader loss cones. 
Large-$R$ distributions have smaller loss cones, but their projections have large positive slopes at low $v_\perp$. 
Low-$R$ distributions have more free energy, but less of that free energy is accessible to flute-like modes. 
Practically speaking, this implies the counterintuitive result that stronger magnets (higher $R$) can make this class of loss-cone instabilities more difficult to stabilize even as they decrease the size of the loss cone. 

From a more academic standpoint, this also provides a new example of a complex plasma phenomenon that can be captured -- at least in part -- by relatively simple thermodynamic arguments. 
There has been substantial recent progress in the field adapting some of these dynamics-agnostic theoretical tools to a wider range of plasma physics applications; the hope, generally, is that this kind of argument can allow physical arguments and intuitions that apply to broad classes of systems \citep{Helander2017ii, Helander2020, Kolmes2020MaxEntropy, Kolmes2022TemperatureScreening, Mackenbach2022, Mackenbach2023Available, Mackenbach2023Miller, Zhdankin2022, Ewart2023}. 

There are two important caveats to point out here. 
The first is that this applies only to flute-like modes. 
Not all instabilities associated with the velocity-space structure in mirror machines are flute-like. 
For example, the Alfv\'en Ion Cyclotron (AIC) mode is electromagnetic and has finite $k_{||}$. 
Even without close analysis, it is straightforward to see that the AIC will not be subject to the same stability conditions that apply to its flute-like cousins. 
Note, in particular, that the AIC can be unstable in bi-Maxwellian distributions (Maxwellian in the perpendicular and parallel directions but with $T_{||} \neq T_\perp$) \citep{Sagdeev1960, Hanson1984, Post1987}. 
The perpendicular projection of a bi-Maxwellian is a monotonically decreasing function of $v_\perp^2$, regardless of the values of $T_{||}$ and $T_\perp$. 
Therefore, the bi-Maxwellian is a ground state against flute-like rearrangements, and would not be unstable against the flute-like modes discussed in the rest of the paper. 
Rotation should still tend to stabilize the AIC mode, since it reduces the anisotropy of the distribution, but the physical picture for that stabilization (and the threshold at which stability is reached) will be different for this and other non-flute-like modes. 

The second caveat is that this analysis is focused on linear stabilization thresholds and does not consider mode saturation. 
In the limit where $R \rightarrow \infty$, the linear theory predicts that the stability thresholds for flute-like loss-cone instabilities become unattainable. 
However, there is some threshold at which the loss cone's size has been diminished so much that the mode cannot access enough free energy to be dangerous: it should saturate at a vanishingly low level in the limit of infinitely large $R$. 

\section*{Acknowledgements}

The authors thank Greta Li, Mikhail Mlodik, Jean-Marcel Rax, and Tal Rubin for helpful conversations. 

\section*{Funding}
This work was supported by ARPA-E Grant No. DE-AR0001554. 
This work was also supported by the DOE Fusion Energy Sciences Postdoctoral Research Program, administered by the Oak Ridge Institute for Science and Education (ORISE) and managed by Oak Ridge Associated Universities (ORAU) under DOE Contract No. DE-SC0014664.

\section*{Declaration of Interests}
The authors report no conflict of interest.

%

\bibliographystyle{jpp}

\bibliography{Master_LC, LC_supplemental}

\end{document}